\documentclass[12pt,a4paper]{article}
\usepackage[T1]{fontenc}
\usepackage[sc,osf]{mathpazo}
\usepackage{a4wide}  
\usepackage{amsfonts, amsthm}
\usepackage{amssymb}
\usepackage{amsmath}
\usepackage{bbm}
\usepackage{mathrsfs}
\usepackage{booktabs} % for much better looking tables
\usepackage{ifpdf}
\ifpdf
\usepackage[pdftex,unicode,implicit]{hyperref}
\hypersetup{%
  pdftitle    = {On 2-dimensional K\"ahler metrics with one holomorphic isometry}
  pdfkeywords = {Kahler, hyper-Kahler, isometry, holomorphic isometry,
    triholomorphic isometry, mono-holomorphic isometry, Gibbons-Hawking metric},
  pdfauthor   = {Samuele Chimento, Tomas Ortin},
  plainpages  = true,
  colorlinks  = true,
  citecolor   = blue,
  urlcolor    = red,
  linkcolor   = black
}
\newcommand{\hepth}[1]{{\tt
\href{http://www.arXiv.org/abs/hep-th/#1}{hep-th/#1}}}

\newcommand{\arxiv}[1]{{\tt arXiv:\href{http://www.arXiv.org/abs/#1}{#1}}}
\newcommand{\doi}[1]{{\tt DOI:\href{http://dx.doi.org/#1}{#1}}}
\newcommand{\euclid}[1]{{\tt Project Euclid:\href{http://projecteuclid.org/euclid.jdg/#1}{#1}}}
\else
  \usepackage[dvips]{graphicx}
  \usepackage[unicode,implicit]{hyperref}
  \newcommand{\hepth}[1]{{\tt hep-th/#1}}

  \newcommand{\arxiv}[1]{{\tt arXiv:#1}}
  \newcommand{\doi}[1]{{\tt DOI:#1}}
\newcommand{\euclid}[1]{{\tt Project Euclid:\href{http://projecteuclid.org/euclid.jdg/#1}{#1}}}
\fi
\makeatletter
\@addtoreset{equation}{section}
\makeatother

\begin{document}

\begin{flushright}
\small
IFT-UAM/CSIC-16-086\\
\texttt{arXiv:1610.02078 [hep-th]}\\
September 19\textsuperscript{th}, 2016\\
\normalsize
\end{flushright}

\vspace{1.5cm}

\begin{center}

  {\Large {\bf On 2-dimensional K\"ahler metrics with one holomorphic
      isometry}}\footnote{\textbf{Important notice:} After the first
    submission of this paper to the arXiv, we have realized that the main
    result presented in it was implicitly contained in Ref.~\cite{lebrun}, as
    explained in Section~2 of Ref.~\cite{kn:Tod}, something neither we nor the
    experts we consulted before the submission were aware of. While this paper
    cannot be published in a regular scientific journal, we think it still can
    be useful for the scientific community.}
 
\vspace{1.5cm}

\renewcommand{\thefootnote}{\alph{footnote}}
{\sl\large  Samuele Chimento}\footnote{E-mail: {\tt Samuele.Chimento [at] csic.es}}
{\sl\large and Tom\'{a}s Ort\'{\i}n}\footnote{E-mail: {\tt Tomas.Ortin [at] csic.es}}

\setcounter{footnote}{0}
\renewcommand{\thefootnote}{\arabic{footnote}}

\vspace{1.5cm}

{\it Instituto de F\'{\i}sica Te\'orica UAM/CSIC\\
C/ Nicol\'as Cabrera, 13--15,  C.U.~Cantoblanco, E-28049 Madrid, Spain}\\ \vspace{0.3cm}

\vspace{1.8cm}

%%%%%%%%%%%%%%%%%%%%%%%%%%%%%%%%%%%%%%%%%%%%%%%%%%%%%%%%%%%%%%%%%%%%%%

{\bf Abstract}

\end{center}

\begin{quotation}
  We show how to write any K\"ahler metric of complex dimension 2 admitting a
  holomorphic isometry as a simple 1-real-function deformation of a
  Gibbons-Hawking metric. Hyper-K\"ahler metrics with a tri-holomorphic
  isometry (Gibbons-Hawking metrics) or with a mono-holomorphic isometry are
  recovered for particular values of the additional function. The new general
  metric can be used as an Ansatz in several interesting physical problems.
\end{quotation}

\newpage
%%%%%%%%%%%%%%%%%%%%%%%%%%%%%%%%%%%%%%%%%%%%%%%%%%%%%%%%%%%%%%%%%%%%%%
%%%%%%%%%%%%%%%%%%%%%%%%%%%%%%%%%%%%%%%%%%%%%%%%%%%%%%%%%%%%%%%%%%%%%%
%%%%%%%%%%%%%%%%%%%%%%%%%%%%%%%%%%%%%%%%%%%%%%%%%%%%%%%%%%%%%%%%%%%%%%
%%%%%%%%%%%%%%%%%%%%%%%%%%%%%%%%%%%%%%%%%%%%%%%%%%%%%%%%%%%%%%%%%%%%%%
\pagestyle{plain}
%%%%%%%%%%%%%%%%%%%%%%%%%%%%%%%%%%%%%%%%%%%%%%%%%%%%%%%%%%%%%%%%%%%%%%
%%%%%%%%%%%%%%%%%%%%%%%%%%%%%%%%%%%%%%%%%%%%%%%%%%%%%%%%%%%%%%%%%%%%%%
%%%%%%%%%%%%%%%%%%%%%%%%%%%%%%%%%%%%%%%%%%%%%%%%%%%%%%%%%%%%%%%%%%%%%%
%%%%%%%%%%%%%%%%%%%%%%%%%%%%%%%%%%%%%%%%%%%%%%%%%%%%%%%%%%%%%%%%%%%%%%
%%%%%%%%%%%%%%%%%%%%%%%%%%%%%%%%%%%%%%%%%%%%%%%%%%%%%%%%%%%%%%%%%%%%%%

%\tableofcontents

\newpage

%%%%%%%%%%%%%%%%%%%%%%%%%%%%%%%%%%%%%%%%%%%%%%%%%%%%%%%%%%%%%%%%%%%%%%
%%%%%%%%%%%%%%%%%%%%%%%%%%%%%%%%%%%%%%%%%%%%%%%%%%%%%%%%%%%%%%%%%%%%%%
%%%%%%%%%%%%%%%%%%%%%%%%%%%%%%%%%%%%%%%%%%%%%%%%%%%%%%%%%%%%%%%%%%%%%%
%%%%%%%%%%%%%%%%%%%%%%%%%%%%%%%%%%%%%%%%%%%%%%%%%%%%%%%%%%%%%%%%%%%%%%
\section*{Introduction}
%%%%%%%%%%%%%%%%%%%%%%%%%%%%%%%%%%%%%%%%%%%%%%%%%%%%%%%%%%%%%%%%%%%%%%
%%%%%%%%%%%%%%%%%%%%%%%%%%%%%%%%%%%%%%%%%%%%%%%%%%%%%%%%%%%%%%%%%%%%%%
%%%%%%%%%%%%%%%%%%%%%%%%%%%%%%%%%%%%%%%%%%%%%%%%%%%%%%%%%%%%%%%%%%%%%%
%%%%%%%%%%%%%%%%%%%%%%%%%%%%%%%%%%%%%%%%%%%%%%%%%%%%%%%%%%%%%%%%%%%%%%

K\"ahler spaces of complex dimension 2 play very important r\^oles in
physics. Of particular interest for us is their occurrence as \textit{base
  spaces} for supersymmetric solutions of minimal Fayet-Iliopoulos (FI)
U(1)--gauged supergravity in 5 and 6 dimensions (see
Refs.~\cite{Gauntlett:2003fk,Cariglia:2004kk,Bellorin:2007yp,Bellorin:2008we}).
In order to give a closed form to these solutions, a closed form for all
2-dimensional K\"ahler metrics would be needed. This is possible if one writes
them as second derivatives of the K\"ahler potential, but then, the
differential equations that determine the rest of the solutions' fields would
be two orders higher and much more difficult to solve.

In the ungauged case one faces a similar problem: finding a closed form for
all the hyper-K\"ahler metric of real dimension 4, and a partial, yet
extremely good, solution is to consider those that admit a triholomorphic
isometry. All these metrics can be written in a simple closed form in terms of
a single real function traditionally denoted by $H$, and are known as
Gibbons-Hawking (GH) metrics
\cite{Gibbons:1979zt,Gibbons:1979xm}. Furthermore, these metrics can be
dimensionally reduced along the isometric direction, establishing fruitful
relations between 5- and 4-dimensional supersymmetric supergravity
solutions. Supersymmetric solutions of 5-dimensional supergravity with a GH
base include (single or multicenter, static and rotating) black-holes and
black rings.

It is, then, natural, to consider K\"ahler metrics admitting one holomorphic
isometry in the U(1) gauged case, but no closed form for them has been given
in the literature. The goal of this paper is to close this gap: we are going
to show how any K\"ahler metric of complex dimension 2 admitting one
holomorphic Killing vector can be written in a simple way in terms of two
independent real functions $H,W$. Particular cases such as GH metrics, or
hyperK\"ahler metrics with mono-holomorphic isometries \cite{Boyer:1982mm,
  gegdas, Cvetic:2001sr} or the scalar--flat K\"ahler metrics with a
holomorphic isometry considered by Lebrun in Ref.~\cite{lebrun} are contained
in this general form and can be recovered by imposing additional conditions on
the function $W$.

%%%%%%%%%%%%%%%%%%%%%%%%%%%%%%%%%%%%%%%%%%%%%%%%%%%%%%%%%%%%%%%%%%%%%%
%%%%%%%%%%%%%%%%%%%%%%%%%%%%%%%%%%%%%%%%%%%%%%%%%%%%%%%%%%%%%%%%%%%%%%
%%%%%%%%%%%%%%%%%%%%%%%%%%%%%%%%%%%%%%%%%%%%%%%%%%%%%%%%%%%%%%%%%%%%%%
%%%%%%%%%%%%%%%%%%%%%%%%%%%%%%%%%%%%%%%%%%%%%%%%%%%%%%%%%%%%%%%%%%%%%%
\section{4-d K\"ahler metrics with one holomorphic isometry}
\label{sec-isok}
%%%%%%%%%%%%%%%%%%%%%%%%%%%%%%%%%%%%%%%%%%%%%%%%%%%%%%%%%%%%%%%%%%%%%%
%%%%%%%%%%%%%%%%%%%%%%%%%%%%%%%%%%%%%%%%%%%%%%%%%%%%%%%%%%%%%%%%%%%%%%
%%%%%%%%%%%%%%%%%%%%%%%%%%%%%%%%%%%%%%%%%%%%%%%%%%%%%%%%%%%%%%%%%%%%%%
%%%%%%%%%%%%%%%%%%%%%%%%%%%%%%%%%%%%%%%%%%%%%%%%%%%%%%%%%%%%%%%%%%%%%%

\textbf{Theorem:} Any K\"ahler metric of real dimension 4 admitting a
holomorphic isometry can be locally written in the form 

\begin{equation}
\label{eq:final_metric} 
ds^{2} 
= 
H^{-1}\left( dz+\chi \right)^{2}
+H\left\{(dx^{2})^{2}+W^{2}(\vec{x})[(dx^{1})^{2}+(dx^{3})^{2}]\right\}\,,
\end{equation} 

\noindent
with the functions $H$ and $W$, and the 1-form $\chi$, depending only on the
three coordinates $x^{i}$, $i=1,2,3$, and satisfying the constraints

\begin{equation}
\label{eq:constraintijcurved}
\begin{array}{rcl}
(d\chi)_{\underline{1}\underline{2}} 
& = &
\partial_{\underline{3}}H\, ,
\\
& & \\
(d\chi)_{\underline{2}\underline{3}} 
& = &
\partial_{\underline{1}}H\, ,
\\
& & \\
(d\chi)_{\underline{3}\underline{1}} 
& = &
\partial_{\underline{2}}\left(W^{2}H\right)\, .
\end{array}
\end{equation}

Conversely, any metric of the above form is a K\"ahler metric admitting a
holomorphic isometry.

\noindent
\textbf{Remark:} The integrability condition of the above three equations is

\begin{equation}
\label{eq:integrability}
\mathfrak{D}^{2}H\equiv \partial_{\underline{1}}\partial_{\underline{1}} H
+\partial_{\underline{2}}\partial_{\underline{2}}\left(W^{2} H\right)
+\partial_{\underline{3}}\partial_{\underline{3}} H 
= 0\, .
\end{equation}

\noindent
Notice that, in general, this equation is not (proportional to) the Laplace
equation in the 3-dimensional metric. The 3-dimensional Laplacian takes the
form

\begin{equation}
\overline{\nabla}^{2}H 
= 
\frac{1}{W^{2}}  
\left[\partial_{\underline{1}}\partial_{\underline{1}} H
+\partial_{\underline{2}}\left(W^{2} \partial_{\underline{2}} H\right)
+\partial_{\underline{3}}\partial_{\underline{3}} H 
 \right]\, ,
\end{equation}

\noindent
and, therefore, the integrability equation is proportional to the Laplace
equation for $x^{2}$-independent conformal factors $W$.

On the other hand, locally, the metric (\ref{eq:final_metric}) is entirely
determined by the two real functions $H$ and $W$. Once a solution $(H,W)$ of
Eq.~(\ref{eq:integrability}) has been found, the 1-form $\chi$ is determined
from (\ref{eq:constraintijcurved}) up to an irrelevant closed 1-form.

\vspace{1cm}

\noindent
\textbf{Proof of the theorem}: Any 4-dimensional Euclidean metric admitting
one isometry can be written in the form

\begin{equation}
\label{eq:4dmetric}
d\hat{s}^{2} 
= 
H^{-1}(dz+\chi)^{2} 
+
H\gamma_{\underline{i}\underline{j}}dx^{i}dx^{j}\, , 
\end{equation}

\noindent
where $z=x^{\sharp}$ is the coordinate adapted to the isometry and where the
3-dimensional function $H$, the 1-form $\chi=\chi_{\underline{i}}dx^{i}$ and
the metric $\gamma_{\underline{i}\underline{j}}dx^{i}dx^{j}$, $i,j=1,2,3$ are
$z$-independent and orthogonal to the Killing vector
$k^{\underline{m}}=\delta_{z}{}^{\underline{m}}$. We denote the coordinate
base indices by $\{\underline{m}\}=\{z,\underline{i}\}$ and the tangent space
indices by $\{ m \}=\{\sharp,i\}$. We will denote 3-dimensional structures
(connection, curvature etc.) by an overline.

A convenient basis of Vierbeins is 

\begin{equation}
\label{eq:4dvierbein}
\left\{
\begin{array}{rcl}
\hat{V}^{\sharp} & = & H^{-1/2}(dz+\chi)\, ,\\
& & \\
\hat{V}^{i} & = & H^{1/2}v^{i}\, ,\\
\end{array}
\right.
\hspace{2cm}
\left\{
\begin{array}{rcl}
\hat{V}_{\sharp} & = & H^{1/2} \partial_{z}\, ,\\
& & \\
\hat{V}_{i} & = & H^{-1/2}(\partial_{i} -\chi_{i}\partial_{z})\, ,
\end{array}
\right.
\end{equation}

\noindent
where $v^{i}= v^{i}{}_{\underline{j}}dx^{j}$ are Dreibeins of the metric
$\gamma_{\underline{i}\underline{j}}$, $\partial_{i}\equiv
v_{i}{}^{\underline{j}}\partial_{\underline{j}}$ and $\chi_{i}\equiv
v_{i}{}^{\underline{j}}\chi_{\underline{j}}$.

The non-vanishing components of the spin connection 1-form, defined through
the structure equation $\mathcal{D}\hat{V}^{m}\equiv d\hat{V}^{m}-\varpi^{m}{}_{n}\wedge
\hat{V}^{n}=0$ are

\begin{equation}
\label{eq:4d_spin_conn}
\begin{array}{rclrcl}
\varpi_{\sharp \sharp i} 
& = & 
\tfrac{1}{2} H^{-3/2}\partial_{i}H\, ,
\hspace{1.5cm}
&
\varpi_{\sharp ij} 
& = & 
\tfrac{1}{2} H^{-3/2} (d\chi)_{ij}\, ,
\\
& & & & & \\
\varpi_{i\sharp j} 
& = & 
\varpi_{\sharp ij}\, , 
&
\varpi_{kij} 
& = & 
H^{-1/2}\overline{\omega}_{kij} 
+H^{-3/2}\partial_{[i}H\delta_{j]k}\, ,
\end{array}
\end{equation}

\noindent
where $(d\chi)_{ij} = 2v_{i}{}^{\underline{k}}
v_{j}{}^{\underline{l}}\partial_{[\underline{j}}\chi_{\underline{l}]}$ and
$\overline{\omega}_{kij}$ is the 3-dimensional connection defined by
$\overline{\mathcal{D}} v^{i}=dv^{i}-\overline{\omega}^{i}{}_{j}\wedge v^{j}=0$.

For the manifold to be K\"ahler, there must exist a globally defined almost
complex structure $J^{\underline{m}}{}_{\underline{n}}$,

\begin{equation}
\label{eq:almostcomplex}
J^{\underline{m}}{}_{\underline{p}}J^{\underline{p}}{}_{\underline{n}}
=
-\delta^{\underline{m}}{}_{\underline{n}}\,,
\end{equation}

\noindent
with respect to which the metric $h_{\underline{m}\underline{n}}$ is
Hermitian,

\begin{equation}
\label{eq:Hermitian}
h_{\underline{m}\underline{n}}
J^{\underline{m}}{}_{\underline{p}}J^{\underline{n}}{}_{\underline{q}}
=
h_{\underline{p}\underline{q}}\, ,
\end{equation}

\noindent
and which is covariantly constant with respect to the Levi-Civita connection,

\begin{equation}
\nabla_{\underline{m}}J^{\underline{n}}{}_{\underline{p}}=0\, .
\end{equation}

\noindent
Eqs.~(\ref{eq:almostcomplex}) and (\ref{eq:Hermitian}) imply that
$J_{\underline{m}\underline{n}}\equiv
h_{\underline{m}\underline{p}}J^{\underline{p}}{}_{\underline{n}}$ is
antisymmetric: \textit{i.e.}~it is a 2-form known as the K\"ahler 2-form. It
is obvious from the covariant constancy of $J$ that the K\"ahler 2-form is
closed. Assuming the other two conditions are met, the closedness of the
K\"ahler 2-form, its covariant constancy or that of the complex structure with
respect to the Levi-Civita connection are equivalent. These statements are
also equivalent to the statement that the holonomy of
$h_{\underline{m}\underline{n}}$ is contained in U$(2)$ and $J$ is the
associated U$(2)$-structure \cite{kn:Joyce}.

In flat four-dimensional indices these conditions are equivalent to 

\begin{eqnarray}
\label{eq:flat_complex_structure}
J_{mn}J_{np}
& = & 
-\delta_{mn}\, ,
\\
& & \nonumber \\
\label{eq:flat_complex_structure_ortho}
J_{mn}
& = &
-J_{nm}\, ,
\\
& & \nonumber \\
\label{eq:flat_complex_structure_conserved}
\nabla_{m}J_{np}
& = &
0\, .
\end{eqnarray}

A $J$ that satisfies the first two conditions and can always be chosen is
given by

\begin{equation}
\label{eq:J_matrix}
(J_{mn}) 
\equiv
\begin{pmatrix}
    \phantom{-}0_{2\times 2} & \phantom{-}\mathbb{1}_{2\times 2}\\
    -\mathbb{1}_{2\times 2}  & \phantom{-} 0_{2\times 2}
   \end{pmatrix}\, .
\end{equation}

We have chosen it to be antiselfdual for the sake of convenience.

Since in this form $J$ is constant, the third condition
Eq.~(\ref{eq:flat_complex_structure_conserved}) is equivalent to the vanishing
of the commutator of $J$ with all the components $m$ of the spin
connection 1-form $\omega_{m}{}^{n}{}_{p}$

\begin{equation}
\label{eq:J_commutator}
\left[\omega_{m},J\right] = 0\, .
\end{equation}

Using the explicit form of the components of the spin connection in
Eqs.~(\ref{eq:4d_spin_conn}), the $m=\sharp$ component of this equation
gives
\begin{eqnarray}
\label{eq:constraint12flat}
\left( d\chi \right)_{12}
& = &
\partial_{3} H\, ,
\\
& & \nonumber \\
\label{eq:constraint23flat}
\left( d\chi \right)_{23}
& = &
\partial_{1} H\, ,
\end{eqnarray}

\noindent
while the $m=i$ components impose the following conditions on the 
components of the spin connection of the 3-dimensional metric
$\gamma_{\underline{i}\underline{j}}$:

\begin{eqnarray}
\label{eq:3d_spin_conn_vanish}
\overline{\omega}_{221}
& = &
\overline{\omega}_{223}=\overline{\omega}_{321}=\overline{\omega}_{123}=0\, ,
\\
& & \nonumber \\
\label{eq:3d_spin_conn_non_vanish}
\overline{\omega}_{112}
& = &
\overline{\omega}_{332}
=
\frac{1}{2H}\left[ \left( d\chi \right)_{13}+\partial_2 H \right]\, .
\end{eqnarray}

(Observe that the components $\overline{\omega}_{113}$, $\overline{\omega}_{213}$
and $\overline{\omega}_{313}$ are not constrained by the K\"ahler condition.)

The last condition that we have to impose on the metric
Eq.~(\ref{eq:4dmetric}) is that the isometry is holomorphic, that is: the
Killing vector $k$ preserves the complex structure

\begin{equation}
\pounds_{k}J=0\, .
\end{equation}

\noindent
However, given the choices made here, this turns out to be automatically true
and does not provide any further conditions. 

We now remind the reader that the condition $\pounds_{k}J=0$ together with the
closedness of the K\"ahler 2-form lead to

\begin{equation}
\pounds_{k}J= i_{k}(dJ) +d(i_{k}J)= d(i_{k}J) =0\, ,  
\end{equation}

\noindent
which implies the existence of a real function $\mathcal{P}$ known as the
\textit{momentum map} such that

\begin{equation}
\label{eq:momentummapdef}
i_{k}J = -d\mathcal{P}\, .
\end{equation}

The conditions (\ref{eq:3d_spin_conn_vanish}) imply that $v^{2}$ is a closed
1-form, $dv^{2}=0$, which means that it is possible to choose a coordinate
$x^{2}$ such that, locally,

\begin{equation}
\label{eq:v2}
v^{2}=dx^{2}\, .
\end{equation}

This can also be seen in a different way: given the form of $J$ in
\eqref{eq:J_matrix}, the K\"ahler form is

\begin{equation}
\mathcal{J}
=
\tfrac{1}{2} J_{mn}\,e^{m}\wedge e^{n} 
= 
e^{\sharp} \wedge e^{2} + e^{1} \wedge e^{3}\, ,
\end{equation}

\noindent
which implies, since $e^{\sharp} = H^{1/2} k_{\underline{m}}dx^{m}$ and $e^{2}
= H^{1/2} v^{2}$,

\begin{equation}
\label{eq:v2_int_der}
v^{2} = \imath_{k}\mathcal{J}\, .
\end{equation}

\noindent
Comparing this equation with Eq.~(\ref{eq:momentummapdef}) we see that

\begin{equation}
v^{2} = -d\mathcal{P}\, ,
\end{equation}

\noindent
and we conclude that we have chosen, as coordinate $x^{2}$, (minus) the
momentum map $x^{2}=-\mathcal{P}$.

Apart form the condition $dv^{2}=0$, the information on the 3-dimensional
metric given by Eqs.~(\ref{eq:3d_spin_conn_vanish}) and
(\ref{eq:3d_spin_conn_non_vanish}) can be summarized in the conditions

\begin{equation}
\label{eq:dv_cond}
dv^{1}\wedge v^{1}=dv^{3}\wedge v^{3}\, ,
\hspace{1.5cm}
dv^{1}\wedge v^{3} = -dv^{3}\wedge v^{1}\, .
\end{equation}

\noindent
Introducing another two coordinates $x^{1,3}$, in general one will have
$v^{1}= v^{1}{}_{\underline{i}} dx^{i}$ and $v^{3} = v^{3}{}_{\underline{i}}
dx^{i}$, with $i=1,2,3$ but the components $v^{1}{}_{\underline{2}}$ and
$v^{3}{}_{\underline{2}}$ can always be set to zero with a coordinate change
$x^{1,3}\rightarrow F^{1,3}(\vec{x})$ such that

\begin{equation}
\partial_{\underline{2}} F^{1}
=
\frac{v^{1}{}_{\underline{3}} v^{3}{}_{\underline{2}}
-v^{1}{}_{\underline{2}} v^{3}{}_{\underline{3}}}
{v^{1}{}_{\underline{1}}v^{3}{}_{\underline{3}}
-v^{3}{}_{\underline{1}} v^{1}{}_{\underline{3}}}\, ,
\hspace{1.5cm}
\partial_{\underline{2}} F^{3}
=
\frac{v^{3}{}_{\underline{1}} v^{1}{}_{\underline{2}}
-v^{3}{}_{\underline{2}} v^{1}{}_{\underline{1}}}
{v^{1}{}_{\underline{1}} v^{3}{}_{\underline{3}}
-v^{3}{}_{\underline{1}} v^{1}{}_{\underline{3}}}\, .
\end{equation}

If $v^{1,3}{}_{\underline{2}}=0$, then Eqs.~(\ref{eq:dv_cond}) imply the
following relations between the Dreibein components and their partial
derivatives with respect to the coordinate $x^{2}$ hold:

\begin{eqnarray}
\partial_{\underline{2}} v^{3}{}_{\underline{1}} 
& = & 
\frac{\partial_{\underline{2}} v^{1}{}_{\underline{1}}
(v^{1}{}_{\underline{1}} v^{1}{}_{\underline{3}}+v^{3}{}_{\underline{1}}
  v^{3}{}_{\underline{3}})
-\partial_{\underline{2}} v^{1}{}_{\underline{3}}
[(v^{1}{}_{\underline{1}})^{2}+(v^{3}{}_{\underline{1}})^{2}]}
{v^{1}{}_{\underline{1}}v^{3}{}_{\underline{3}}
-v^{3}{}_{\underline{1}} v^{1}{}_{\underline{3}}}\, , 
\nonumber\\
& & \label{eq:der2v3}\\
\partial_{\underline{2}} v^{3}{}_{\underline{3}} 
& = &
\frac{\partial_{\underline{2}} v^{1}{}_{\underline{1}}
[(v^{1}{}_{\underline{1}})^{2}+(v^{3}{}_{\underline{1}})^{2})
-\partial_{\underline{2}} v^{1}{}_{\underline{3}}
(v^{1}{}_{\underline{1}} v^{1}{}_{\underline{3}}+v^{3}{}_{\underline{1}}
v^{3}{}_{\underline{3}})}
{v^{1}{}_{\underline{1}}v^{3}{}_{\underline{3}}
-v^{3}{}_{\underline{1}} v^{1}{}_{\underline{3}}}\, .
\nonumber
\end{eqnarray}

For a fixed value of $x^{2}$ (and treating it as a constant) there always
exists a coordinate change $x^{1,3}\rightarrow G^{1,3}(x^{1},x^{3})$ allowing
to rewrite the 2-dimensional metric $ds_{2}^{2} = (v^{1})^{2}+(v^{3})^{2}$ in
conformally flat form $ds_{2}^{2}=W^{2}(\vec{x})[(dx^{1})^{2}+(dx^{3})^{2}]$.
The derivatives of the functions $G^{1,3}$ that do the trick satisfy the
conditions

\begin{equation}
\partial_{\underline{1}}G^{3}
= 
A\, \partial_{\underline{1} }G^{1} +B\, \partial_{\underline{3} }G^{1}\, ,
\hspace{1cm} 
\partial_{\underline{3}} G^{3}
= 
A\, \partial_{\underline{3} }G^{1} - B\, \partial_{\underline{1} }G^{1}\, ,
\end{equation}

\noindent
with

\begin{equation}
A
=
-\frac{v^{1}{}_{\underline{1}}v^{1}{}_{\underline{3}}
+v^{3}{}_{\underline{1}}v^{3}{}_{\underline{3}}}
{(v^{1}{}_{\underline{3}})^{2}+(v^{3}{}_{\underline{3}})^{2}}\, ,
\hspace{1cm}
B
=
\pm\frac{v^{1}{}_{\underline{3}}v^{3}{}_{\underline{1}}
-v^{1}{}_{\underline{1}}v^{3}{}_{\underline{3}}}
{(v^{1}{}_{\underline{3}})^{2}+(v^{3}{}_{\underline{3}})^{2}}\, .
\end{equation}

In general, if the non-vanishing components of the Dreibein depend on $x^{2}$,
the functions $A$ and $B$ depend on $x^{2}$, and the functions $G^{1,3}$
cannot satisfy the above equations being independent of $x^{2}$. On the other
hand, if $G^{1,3}$ depended on $x^{2}$ the above equations would not make
sense as we would have to include partial derivatives with respect to
$x^{2}$. 

In the present case, however, it turns out that Eqs.~(\ref{eq:der2v3}) imply
that $\partial_{\underline{2}} A=\partial_{\underline{2}} B=0$, guaranteeing
that the same coordinate change with $x^{2}$-independent $G^{1,3}$ allows to
write the 2-dimensional metric in conformally flat form even if the components
of $v^{1,3}$ depend on the third coordinate $x^{2}$.

We conclude that we can always choose coordinates in the 2-dimensional metric
such that the non-trivial Dreibein are given by 

\begin{equation}
\label{eq:vtilde} 
v^{1,3} = W(\vec{x}) dx^{1,3}\, ,
\end{equation} 

\noindent
and the 3-dimensional metric is

\begin{equation}
d\overline{s}^{2} 
= 
\gamma_{\underline{i}\underline{j}}dx^{i}dx^{j}  
=
(dx^{2})^{2}+W^{2}(\vec{x})[(dx^{1})^{2}+(dx^{3})^{2}]\, . 
\end{equation}

Computing explicitly the spin connection components of this metric 
in terms of $W^2$, the constraint Eq.~(\ref{eq:3d_spin_conn_non_vanish}) becomes

\begin{equation}
\label{eq:constraint13flat} 
\left( d\chi \right)_{31}
=
\partial_{2} H
+H\partial_{2}\log{W^{2}}\, ,
\end{equation}

\noindent
which is the third condition in Eqs.~(\ref{eq:constraintijcurved}), proving
the first part of the theorem.

Showing that the inverse is also true, that is, that any metric of the form
Eq.~(\ref{eq:final_metric}) satisfying the constraints
Eqs.~(\ref{eq:constraintijcurved}) is K\"ahler, is straightforward. One can
introduce a Dreibein $v^{i}$ given by Eqs.~(\ref{eq:v2}) and
(\ref{eq:vtilde}), a Vierbein as in Eqs.~(\ref{eq:4dvierbein}) and a complex
structure as in Eq.~(\ref{eq:J_matrix}). Equations
(\ref{eq:flat_complex_structure}) and (\ref{eq:flat_complex_structure_ortho})
are automatically satisfied, and using the constraints
(\ref{eq:constraintijcurved})) it is easy to verify that
Eq.~(\ref{eq:flat_complex_structure_conserved}) which is again equivalent to
Eq.~(\ref{eq:J_commutator}), is also satisfied.

\vspace{.5cm}
\textit{Q.E.D.}
\vspace{.5cm}

In the preceding discussion we have ignored the existence of a K\"ahler
potential. Finding the K\"ahler potential from the metric in a
given set of real coordinates is not an easy task. Observe, however, that the
main equation that the functions that define our metric $H,W$ satisfy,
Eq.~(\ref{eq:integrability}), can always be solved by introducing 
a real function $\mathcal{K}(x^{1},x^{2},x^{3})$ and defining

\begin{equation}
H\equiv \partial_{\underline{2}}^{2} \mathcal{K}\, ,
\hspace{1cm}
W^{2} \equiv -H^{-1} 
\left(\partial_{\underline{1}}^{2}+\partial_{\underline{3}}^{2} \right)\mathcal{K}\, .
\end{equation}

The components $\chi_{\underline{1}},\chi_{\underline{3}}$ of the 1-form
$\chi$ satisfying Eq.~(\ref{eq:constraintijcurved}) can also be derived from
$\mathcal{K}$, as long as we choose coordinates such that
$\chi_{\underline{2}}=0$. They are given by

\begin{equation}
\chi_{\underline{1}} = -\partial_{\underline{3}}\partial_{\underline{2}}
\mathcal{K}\, ,
\hspace{1cm}
\chi_{\underline{3}} = \partial_{\underline{2}}\partial_{\underline{1}}
\mathcal{K}\, .
\end{equation}

It is tempting to identify $\mathcal{K}$ with the K\"ahler potential. However,
although this is likely to be the case, we have not proven its existence nor
we have proven that the above relations are the unique way of solving the
equations that define the metric. Nevertheless, we can always consider metrics
constructed in this way since they are automatically K\"ahler metrics with a
holomorphic isometry.

%%%%%%%%%%%%%%%%%%%%%%%%%%%%%%%%%%%%%%%%%%%%%%%%%%%%%%%%%%%%%%%%%%%%%%
%%%%%%%%%%%%%%%%%%%%%%%%%%%%%%%%%%%%%%%%%%%%%%%%%%%%%%%%%%%%%%%%%%%%%%
%%%%%%%%%%%%%%%%%%%%%%%%%%%%%%%%%%%%%%%%%%%%%%%%%%%%%%%%%%%%%%%%%%%%%%
%%%%%%%%%%%%%%%%%%%%%%%%%%%%%%%%%%%%%%%%%%%%%%%%%%%%%%%%%%%%%%%%%%%%%%
\section{Special cases}
\label{sec-special}
%%%%%%%%%%%%%%%%%%%%%%%%%%%%%%%%%%%%%%%%%%%%%%%%%%%%%%%%%%%%%%%%%%%%%%
%%%%%%%%%%%%%%%%%%%%%%%%%%%%%%%%%%%%%%%%%%%%%%%%%%%%%%%%%%%%%%%%%%%%%%
%%%%%%%%%%%%%%%%%%%%%%%%%%%%%%%%%%%%%%%%%%%%%%%%%%%%%%%%%%%%%%%%%%%%%%
%%%%%%%%%%%%%%%%%%%%%%%%%%%%%%%%%%%%%%%%%%%%%%%%%%%%%%%%%%%%%%%%%%%%%%

The scalar curvature of the metric (\ref{eq:final_metric}) can be written in 
the compact form

\begin{equation}
 \hat{R} = \hat{\nabla}^{2}\log{W^{2}}=H^{-1}\overline{\nabla}^{2}\log{W^{2}}\,,
\end{equation}

\noindent
where $\hat{\nabla}^{2}$ is the 4-dimensional Laplacian operator. If one were to 
impose the requirement of scalar-flatness on the metric, this would thus 
translate to an equation for $W^{2}$ which is known in the physics literature
as the SU$(\infty)$ or 3D Toda equation:

\begin{equation}
\label{eq:toda}
 \left(\partial_{\underline{1}}^{2}+\partial_{\underline{3}}^{2}
 \right)\nu+\partial_{\underline{2}}^{2}\, e^{\nu}=0\,,
\end{equation}

\noindent
with $\nu\equiv\log{W^{2}}$. In this case our result reduces to the one of
LeBrun \cite{lebrun}, which however was obtained imposing from the beginning a
vanishing scalar curvature.

It is always possible to introduce two additional complex structures $J^{(2,3)}$
satisfying together with $J^{(1)}\equiv J$ the unit quaternionic algebra

\begin{equation} 
J^{(x)} J^{(y)} 
= 
-\delta^{xy} \mathbb{1} +\epsilon^{xyz}J^{(z)}\,.
\end{equation}
In particular one can choose them to be of the form
\begin{equation}
\label{eq:J23_matrices} 
J^{(2)}
=
\begin{pmatrix} 
i\sigma_{2} & \phantom{-i}0_{2\times 2}\\ 
\phantom{i}0_{2\times 2} & - i\sigma_{2} \\
\end{pmatrix}\, ,
\hspace{1.5cm}
J^{(3)}
=
\begin{pmatrix} 
\phantom{-i}0_{2\times 2} & -i\sigma_{2} \\ 
- i\sigma_{2} & \phantom{-i}0_{2\times 2} \\
\end{pmatrix}\, .
\end{equation} 

\noindent
Observe that, with this choice, the 1-forms $v^{1}$ and $v^{3}$ can be written
in terms of these complex structures in a similar way as $v^{2}$ in
\eqref{eq:v2_int_der}, namely

\begin{equation} 
v^{1} 
= 
\imath_{k}J^{(2)}\, ,
\hspace{1.5cm}
v^{3} 
=
-\imath_{k}J^{(3)}\, .
\end{equation} 

\noindent
Of course in general these complex structures are not covariantly constant, in fact one has

\begin{eqnarray}
\label{eq:df2=pf3}
\hat{\nabla}_{m}J^{(2)}{}_{np} 
& = & 
\hat{P}_{m}J^{(3)}{}_{np}\, ,
\\
& & \nonumber \\
\label{eq:df3=-pf2}
\hat{\nabla}_{m}J^{(3)}{}_{np} 
& = & 
-\hat{P}_{m}J^{(2)}{}_{np}\, ,
\end{eqnarray}

\noindent
with the components of the 1-form $P$ in (4-dimensional) flat indices 
given by

\begin{equation}
\label{eq:P_vector} 
\hat{P}_{m} = \hat{J}_{m}{}^{n}\,\partial_{n}\log{W}\, .
\end{equation}

\noindent
Actually, the most general possible form for $J^{(2,3)}$ would be 

\begin{equation}
\label{eq:gen_J23}
 J^{(2)\,\prime} =\cos{\theta}\, J^{(2)}+\sin{\theta}\, J^{(3)}\,,\qquad 
 J^{(3)\, \prime} = \cos{\theta}\, J^{(3)}-\sin{\theta}\,J^{(2)}\,,
\end{equation}

\noindent
for some function $\theta$, in which case

\begin{equation}
 \hat{P}^\prime= \hat{P} -d\theta\,.
\end{equation}

If one chooses $H=\partial_{\underline{2}}\log W^{2}$, then the integrability 
condition (\ref{eq:integrability}) reduces to the derivative with respect to
$x^{2}$ of the Toda equation (\ref{eq:toda}). Therefore it is automatically
satisfied if one imposes Eq. (\ref{eq:toda}), which as we have seen is 
equivalent to the requirement of scalar-flatness. In this case one gets

\begin{equation}
 \chi = \partial_{\underline{1}}\log W^{2} dx^3 - \partial_{\underline{3}}\log W^{2} dx^1
\end{equation}

\noindent
and 

\begin{equation}
 \hat{P}=\tfrac12 dz\,,
\end{equation}

\noindent
which means that the complex structures given by (\ref{eq:gen_J23}) with $\theta=z/2$
are covariantly constant and the space is hyperK\"ahler, while not being preserved by 
the isometry. These hyperK\"ahler metrics with a mono-holomorphic isometry
were studied in \cite{Boyer:1982mm, gegdas, Cvetic:2001sr}.

If instead $W$ is taken to be constant, the 3-dimensional metric is flat and the
constraint Eqs.~(\ref{eq:constraintijcurved}) reduce to

\begin{equation} 
d\chi=\star_{3} dH\, ,
\end{equation} 

\noindent
which implies that $H$ is harmonic. The 1-form $\hat P$ vanishes, which means
that $J^{(2,3)}$ are covariantly constant. In this case they are also
preserved by the isometry, $\pounds_{k}J^{(2,3)}=0$. The metric
Eq.~(\ref{eq:final_metric}) is, then, a Gibbons-Hawking metric
\cite{Gibbons:1979zt,Gibbons:1979xm}. Therefore, it is a hyperK\"ahler metric
admitting a triholomorphic isometry. In this scheme, the non-triviality of the
conformal factor $W$ can be seen as the obstruction for the K\"ahler metric
with a holomorphic isometry to be a hyper-K\"ahler metric with a
triholomorphic isometry.

%%%%%%%%%%%%%%%%%%%%%%%%%%%%%%%%%%%%%%%%%%%%%%%%%%%%%%%%%%%%%%%%%%%%%%
%%%%%%%%%%%%%%%%%%%%%%%%%%%%%%%%%%%%%%%%%%%%%%%%%%%%%%%%%%%%%%%%%%%%%%
%%%%%%%%%%%%%%%%%%%%%%%%%%%%%%%%%%%%%%%%%%%%%%%%%%%%%%%%%%%%%%%%%%%%%%
%%%%%%%%%%%%%%%%%%%%%%%%%%%%%%%%%%%%%%%%%%%%%%%%%%%%%%%%%%%%%%%%%%%%%%
\section{An example}
\label{sec-example}
%%%%%%%%%%%%%%%%%%%%%%%%%%%%%%%%%%%%%%%%%%%%%%%%%%%%%%%%%%%%%%%%%%%%%%
%%%%%%%%%%%%%%%%%%%%%%%%%%%%%%%%%%%%%%%%%%%%%%%%%%%%%%%%%%%%%%%%%%%%%%
%%%%%%%%%%%%%%%%%%%%%%%%%%%%%%%%%%%%%%%%%%%%%%%%%%%%%%%%%%%%%%%%%%%%%%
%%%%%%%%%%%%%%%%%%%%%%%%%%%%%%%%%%%%%%%%%%%%%%%%%%%%%%%%%%%%%%%%%%%%%%

A non-trivial example of K\"ahler manifold admitting one isometry is the
non-compact symmetric space $\overline{\mathbb{CP}}^{2}=$SU$(1,2)/$U$(2)$.
In supergravity it arises as the base space of AdS$_{5}$, which can be
constructed as a U$(1)$ bundle over $\overline{\mathbb{CP}}^{2}$
\cite{Gibbons:2011sg}\footnote{This is the non-compact version of the Hopf
  fibrations studied by Trautman in Ref.~\cite{Trautman:1977im}.}. 
Its metric is usually given in terms of complex coordinates $\zeta^i$, 
$i=1,2$, as

\begin{equation}
\mathcal{G}_{ij^{*}}
=
\frac{ \delta_{ij^{*}}}{1-\zeta^{k}\zeta^{*\, k^{*}}}
+
  \frac{\zeta^{*\, i^{*}}\zeta^{j}}{(1-\zeta^{k}\zeta^{*\, k^{*}})^{2}}\,.  
\end{equation}

Introducing the real coordinates

\begin{equation}
\zeta^{1} 
= \tanh{\rho}\cos{\tfrac{\theta}{2}}\, e^{-\frac{i}{2}(\psi+\varphi)}\, ,
\hspace{1.5cm} 
\zeta^{2} 
= 
\tanh{\rho}\sin{\tfrac{\theta}{2}}\,e^{-\frac{i}{2}(\psi-\varphi)}\, ,
\end{equation}

\noindent
the line element of $\overline{\mathbb{CP}}^{2}$ takes the form

\begin{equation}
ds^{2} 
= 
d\rho^{2} +\tfrac{1}{4} \sinh^{2}{\rho}
\left[d\theta^{2}+\sin^{2}{\theta} d\varphi^{2}
+\cosh^{2}{\rho}\left( d\psi+\cos{\theta} d\varphi \right)^{2}\right]\, ,
\end{equation}
and with the further coordinate change
\begin{equation}
z = \psi\, ,
\hspace{1cm} 
x^{2} =\tfrac{1}{4} \sinh^{2}{\rho}\, ,
\hspace{1cm} 
x^{1} = \tan{\tfrac{\theta}{2}}\cos{\varphi}\, ,
\hspace{1cm} 
x^{3} = \tan{\tfrac{\theta}{2}}\sin{\varphi}\, ,
\end{equation}
it can be brought to the form (\ref{eq:final_metric}), with the functions 
$H$, $W^{2}$ and 1-form $\chi$ that define it given by
\begin{equation}
  \begin{array}{rcl}
H^{-1} 
& = & 
x^{2}( 1+4x^{2})\, ,
\\
& & \\
W^{2} 
& = &
{\displaystyle
\frac{4x^{2}}{H[1+(x^{1})^{2}+(x^{3})^{2}]^{2}}\, ,
}
\\
& & \\
\chi 
& = & 
{\displaystyle
\frac{[1-(x^{1})^{2}-(x^{3})^{2}]}{[1+(x^{1})^{2}+(x^{3})^{2}]}
\frac{x^{1}dx^{3}-x^{3}dx^{1}}{(x^{1})^{2}+(x^{3})^{2}}\, .
}
\end{array}
\end{equation}

The functions $W$ and $H$ for this metric have been given in Ref.~

%%%%%%%%%%%%%%%%%%%%%%%%%%%%%%%%%%%%%%%%%%%%%%%%%%%%%%%%%%%%%%%%%%%%%%
%%%%%%%%%%%%%%%%%%%%%%%%%%%%%%%%%%%%%%%%%%%%%%%%%%%%%%%%%%%%%%%%%%%%%%
%%%%%%%%%%%%%%%%%%%%%%%%%%%%%%%%%%%%%%%%%%%%%%%%%%%%%%%%%%%%%%%%%%%%%%
%%%%%%%%%%%%%%%%%%%%%%%%%%%%%%%%%%%%%%%%%%%%%%%%%%%%%%%%%%%%%%%%%%%%%%
\section{Conclusions}
\label{sec-conclusions}
%%%%%%%%%%%%%%%%%%%%%%%%%%%%%%%%%%%%%%%%%%%%%%%%%%%%%%%%%%%%%%%%%%%%%%
%%%%%%%%%%%%%%%%%%%%%%%%%%%%%%%%%%%%%%%%%%%%%%%%%%%%%%%%%%%%%%%%%%%%%%
%%%%%%%%%%%%%%%%%%%%%%%%%%%%%%%%%%%%%%%%%%%%%%%%%%%%%%%%%%%%%%%%%%%%%%
%%%%%%%%%%%%%%%%%%%%%%%%%%%%%%%%%%%%%%%%%%%%%%%%%%%%%%%%%%%%%%%%%%%%%%

With the result we have just proven, the conditions that determine the fields
of supersymmetric solutions of FI-U$(1)$-gauged minimal supergravity in 5 and
6 dimensions must become a set of partial differential equations on a set of
real functions, just as in the ungauged case, although here we expect the
equations to be coupled and non-linear. Still making use of the Ansatz in
Eq.~(\ref{eq:final_metric}), (\ref{eq:constraintijcurved}) should simplify
considerably the problem. Work in this direction is in progress
\cite{kn:CO,kn:CCO}.

%%%%%%%%%%%%%%%%%%%%%%%%%%%%%%%%%%%%%%%%%%%%%%%%%%%%%%%%%%%%%%%%%%%%%%
%%%%%%%%%%%%%%%%%%%%%%%%%%%%%%%%%%%%%%%%%%%%%%%%%%%%%%%%%%%%%%%%%%%%%%
%%%%%%%%%%%%%%%%%%%%%%%%%%%%%%%%%%%%%%%%%%%%%%%%%%%%%%%%%%%%%%%%%%%%%%
%%%%%%%%%%%%%%%%%%%%%%%%%%%%%%%%%%%%%%%%%%%%%%%%%%%%%%%%%%%%%%%%%%%%%%
\section*{Acknowledgments}
%%%%%%%%%%%%%%%%%%%%%%%%%%%%%%%%%%%%%%%%%%%%%%%%%%%%%%%%%%%%%%%%%%%%%%
%%%%%%%%%%%%%%%%%%%%%%%%%%%%%%%%%%%%%%%%%%%%%%%%%%%%%%%%%%%%%%%%%%%%%%
%%%%%%%%%%%%%%%%%%%%%%%%%%%%%%%%%%%%%%%%%%%%%%%%%%%%%%%%%%%%%%%%%%%%%%
%%%%%%%%%%%%%%%%%%%%%%%%%%%%%%%%%%%%%%%%%%%%%%%%%%%%%%%%%%%%%%%%%%%%%%

The authors would like to thank Iosif Bena, Nikolay Bobev, Dietmar Klemm and
Patrick Meessen for interesting conversations and Gary Gibbons for pointing us
to several relevant references.  This work has been supported in part by the
Spanish Ministry of Science and Education grants FPA2012-35043-C02-01 and
FPA2015-66793-P, the Centro de Excelencia Severo Ochoa Program grant
SEV-2012-0249, and the Spanish Consolider-Ingenio 2010 program CPAN
CSD2007-00042. TO wishes to thank M.M.~Fern\'andez for her permanent support.

%%%%%%%%%%%%%%%%%%%%%%%%%%%%%%%%%%%%%%%%%%%%%%%%%%%%%%%%%%%%%%%%%%%%%%
%%%%%%%%%%%%%%%%%%%%%%%%%%%%%%%%%%%%%%%%%%%%%%%%%%%%%%%%%%%%%%%%%%%%%%
%%%%%%%%%%%%%%%%%%%%%%%%%%%%%%%%%%%%%%%%%%%%%%%%%%%%%%%%%%%%%%%%%%%%%%
%%%%%%%%%%%%%%%%%%%%%%%%%%%%%%%%%%%%%%%%%%%%%%%%%%%%%%%%%%%%%%%%%%%%%%
%%%%%%%%%%%%%%%%%%%%%%%%%%%%%%%%%%%%%%%%%%%%%%%%%%%%%%%%%%%%%%%%%%%%%%
%%%%%%%%%%%%%%%%%%%%%%%%%%%%%%%%%%%%%%%%%%%%%%%%%%%%%%%%%%%%%%%%%%%%%%


\begin{thebibliography}{99}

%\cite{Gauntlett:2003fk}
\bibitem{Gauntlett:2003fk}
J.~P.~Gauntlett and J.~B.~Gutowski,
``All supersymmetric solutions of minimal gauged supergravity in five-dimensions,''
Phys.\ Rev.\ D {\bf 68} (2003) 105009.
Erratum: [Phys.\ Rev.\ D {\bf 70} (2004) 089901].
\doi{10.1103/PhysRevD.70.089901}, \doi{10.1103/PhysRevD.68.105009}.
[\hepth{0304064}].
%%CITATION = doi:10.1103/PhysRevD.70.089901, 10.1103/PhysRevD.68.105009;%%

%\cite{Cariglia:2004kk}
\bibitem{Cariglia:2004kk}
M.~Cariglia and O.~A.~P.~Mac Conamhna,
``The General form of supersymmetric solutions of N=(1,0) U(1) and SU(2) gauged supergravities in six-dimensions,''
Class.\ Quant.\ Grav.\  {\bf 21} (2004) 3171
\doi{10.1088/0264-9381/21/13/006}.
[\hepth{0402055}].
%%CITATION = doi:10.1088/0264-9381/21/13/006;%%

%\cite{Bellorin:2007yp}
\bibitem{Bellorin:2007yp}
J.~Bellor\'{\i}n and T.~Ort\'{\i}n,
``Characterization of all the supersymmetric solutions of gauged N=1, d=5 supergravity,''
JHEP {\bf 0708} (2007) 096
\doi{10.1088/1126-6708/2007/08/096}.
[\arxiv{0705.2567} [hep-th]].
%%CITATION = doi:10.1088/1126-6708/2007/08/096;%%

%\cite{Bellorin:2008we}
\bibitem{Bellorin:2008we}
J.~Bellor\'{\i}n,
``Supersymmetric solutions of gauged five-dimensional supergravity with general matter couplings,''
Class.\ Quant.\ Grav.\  {\bf 26} (2009) 195012.
\doi{10.1088/0264-9381/26/19/195012}.
[\arxiv{0810.0527} [hep-th]].
%%CITATION = doi:10.1088/0264-9381/26/19/195012;%%

%\cite{Gibbons:1979zt}
\bibitem{Gibbons:1979zt}
G.~W.~Gibbons and S.~W.~Hawking,
``Gravitational Multi - Instantons,''
Phys.\ Lett.\ B {\bf 78} (1978) 430.
\doi{10.1016/0370-2693(78)90478-1}.
%%CITATION = doi:10.1016/0370-2693(78)90478-1;%%

%\cite{Gibbons:1979xm}
\bibitem{Gibbons:1979xm}
G.~W.~Gibbons and S.~W.~Hawking,
``Classification of Gravitational Instanton Symmetries,''
Commun.\ Math.\ Phys.\  {\bf 66} (1979) 291.
\doi{10.1007/BF01197189}.
%%CITATION = doi:10.1007/BF01197189;%%


%\cite{Boyer:1982mm}
\bibitem{Boyer:1982mm}
C.~P.~Boyer and J.~D.~Finley, III,
``Killing Vectors in Selfdual, Euclidean Einstein Spaces,''
J.\ Math.\ Phys.\  {\bf 23} (1982) 1126.
\doi{10.1063/1.525479}
%%CITATION = doi:10.1063/1.525479;%%

\bibitem{gegdas}
J.~D.~Gegenberg and A.~Das,
``Stationary Riemannian space-times with self-dual curvature,''
Gen.\ Rel.\ and\ Gravitation\ {\bf 16} (1984) 817.
\doi{10.1007/BF00762935}

%\cite{Cvetic:2001sr}
\bibitem{Cvetic:2001sr}
M.~Cvetic, G.~W.~Gibbons, H.~Lu and C.~N.~Pope,
``Orientifolds and slumps in G(2) and spin(7) metrics,''
Annals Phys.\  {\bf 310} (2004) 265
\doi{10.1016/j.aop.2003.10.004}
[\hepth{0111096}].
%%CITATION = doi:10.1016/j.aop.2003.10.004;%%

\bibitem{lebrun}
C.~LeBrun,
``Explicit self-dual metrics on $\mathbb{CP}_2 \# \cdots\#\mathbb{CP}_2$,''
J.\ Diff.\ Geom.\  {\bf 34} (1991) 223.
\euclid{1214446999}.

\bibitem{kn:Joyce}
D.~Joyce,
``Compact Manifolds with Special Holonomy,''
Oxford University Press, U.K. (2000)

%\cite{Gibbons:2011sg}
\bibitem{Gibbons:2011sg}
G.~W.~Gibbons,
``Anti-de-Sitter spacetime and its uses,''
\arxiv{1110.1206} [hep-th].
%%CITATION = ARXIV:1110.1206;%%

%\cite{Trautman:1977im}
\bibitem{Trautman:1977im}
A.~Trautman,
``Solutions of the Maxwell and Yang-Mills Equations Associated with Hopf Fibrings,''
Int.\ J.\ Theor.\ Phys.\  {\bf 16} (1977) 561.
\doi{10.1007/BF01811088}
%%CITATION = doi:10.1007/BF01811088;%%

%\cite{Dunajski:2013qc}
\bibitem{Dunajski:2013qc}
M.~Dunajski, J.~Gutowski and W.~Sabra,
``Enhanced Euclidean supersymmetry, 11D supergravity and $SU(\infty)$ Toda equation,''
JHEP {\bf 1310} (2013) 089.
\doi{10.1007/JHEP10(2013)089}.
[\arxiv{1301.1896} [hep-th]].
%%CITATION = doi:10.1007/JHEP10(2013)089;%%

\bibitem{kn:Tod}
K.P.~Tod,
``Scalar-flat K\"ahler and hyper-K\"ahler metrics from Painlev\'e-III''
Class.\ Quant.\ Grav.\ {\bf 12} (1995) 1535 .
\doi{10.1088/0264-9381/12/6/018}.

\bibitem{kn:CO}
S.~Chimento and T.~Ort\'{\i}n,
``On timelike supersymmetric solutions of gauged minimal 5-dimensional
supergravity'',
to be submitted.

\bibitem{kn:CCO}
P.~Cano, S.~Chimento and T.~Ort\'{\i}n, work in progress.

\end{thebibliography}
\end{document}